\documentclass[twocolumn,pre,showpacs,preprintnumbers]{revtex4}
\usepackage{graphicx}

\begin{document}
\title{Avalanche of Bifurcations and Hysteresis in a 
       Model of Cellular Differentiation} 
\author{G\'abor F\'ath$^{\dag\ddag}$\cite{byline1} and 
        Zbigniew Doma\'nski$^{\ddag}$}
\address{\dag Cavendish Laboratory, University of Cambridge, Madingley
        Road, Cambridge CB3 0HE, England\\
        \ddag Institute of Theoretical Physics, University of Lausanne, 
        CH-1015 Lausanne, Switzerland}
\date{6 May 1999}

\begin{abstract} 
Cellular differentiation in a developping organism is studied 
via a discrete bistable reaction-diffusion model. 
A system of undifferentiated cells is allowed to receive an 
inductive signal emenating from its environment. Depending on the
form of the nonlinear reaction kinetics, this signal can trigger
a series of bifurcations in the system. Differentiation starts
at the surface where the signal is received, and cells change type up to a given
distance, or under other conditions, the differentiation process 
propagates through the whole domain.
When the signal diminishes hysteresis is observed.
\end{abstract} 
\preprint{Physical Review E 60 (4), 4604 (1999)}
\maketitle
%
\section{Introduction}

An adult higher organism, like a human, has some hundreds of
functionally different cell types. The genetic code
stored by the DNA in the cell nucleus is identical in these
cells. This potential information, however, is not utilized completely
by the cells as many genes stay in a dormant, unexpressed state.
The spectrum of genes which are expressed and functioning varies 
from cell type to cell type. 
One of the most fascinating questions in modern biology is how a certain 
cell or a group of cells finds its place and special task (cell type)
in a developing organism.\cite{Gilb} 

From the point of view of dynamical systems, different cell types in the
organism can be associated with different attractors of the common
nonlinear internal dynamics of the cells.\cite{Kauf} The number of 
possible attractors depends on the complexity of this dynamics and on 
the number of genes involved.
It is widely believed that morphogenesis is a precise, 
well-controled series of bifurcations which happen in the proliferating 
and migrating population of cells, generating an ever increasing 
complexity of patterns of differentiated cell regions.\cite{Kauf}

Disregarding some early asymmetric clevages, and non-uniform distribution
of citoplasmic factors in the fertilized egg, cell divisions usually 
produce equivalent daughter cells. Consequently, cell proliferation
leads to an increasing domain of identical cells, where all the 
system parameters are distributed uniformly. Such a subsystem of identical
cells is, however, embedded in, and communicates with other, eventually
already differentiated, groups of cells.

There are essentially two ways that a subsystem of identical cells can 
later differentiate, either as a whole, or in parts: 
(i) There is a critical number of cells above which the spatially 
homogeneous attractor loses stability (Turing instability), leading 
to {\em spontaneous} spatial patterning.\cite{Turi} It is the {\em size} 
(the number of cells) of the increasing domain that plays the role of a
bifurcation parameter. (ii) There is an external {\em inductive} 
signal, emenating from an other group of already differentiated cells, 
which acts as a bifurcation parameter, and drives the system into the new,
spatially inhomogeneous state. In this latter case, timing 
of the signal is crutial.

Many biological examples could be mentioned for the two mechanisms of 
differentiation. Size-driven instabilities [case (i)] take place, e.g.,
in early insect development.\cite{Gilb} A well-studied case is the syncytial
blastoderm stage of the fruit fly {\em Drosophila}, where a
series of patterns of gene expression arise, forming various stripes 
of high and low concentration regions of gene products along the 
anterior-posterior axis. 

Inductive differentiation [case (ii)], on the other hand, is typical 
in later stages of development.\cite{Gilb} As examples, we can mention the 
mesoderm and notochord induction in vertebrates,\cite{Gilb} the
vulva formation in the soil nematode {\em Caenorhabditis 
elegans},\cite{Celegans} and the development of the retina of 
the {\em Drosophila} fly,\cite{retina} where 
inductive influence of the environment cells were clearly demonstrated.

Our aim in this paper is to study a simple example of inductive 
differentiation. 
Emphasis will be put on the aspects of cellular {\em discreteness\/}. The
fact that interacting cells are discrete objects is usually overlooked
in modelling biological pattern formation processes. However, as will be
demonstarted here, spatial discreteness is a source of a variety of
new phenomena with possible biological significance.

\section{The model}

In the following we consider a semi-infinite one-dimensional
chain of cells where the cell distance (lattice constant) is set to
unity. We suppose that each cell in
this system is characterized by the concentration of a single chemical
(the morphogene) whose value (low or high) informs us about the actual
state (type) of the cell. The morphogene concentration in cell $n$ at
time $t$ will be denoted by $u_n(t)$. In an obviously highly 
oversimplified setup the complicated cell biochemistry is reduced to
an effective nonlinear autochatalitic reaction involving the morphogene.
We also assume that the morhogene is diffusive and that the differentiation
process can be described on a reaction-diffusion basis.
Since the cells are discrete objects their diffusive coupling is
modelled by a {\em discrete} Laplacian, and as it will be demonstrated, this 
has far-reaching consequences. The analysis in the following can be
readily generalized to two- or three-dimensional domains with a straight
surface if fluctuations in $u({\bf n},t$ parallel with
the surface can be neglected.

Inside the bulk of the system $2\le n<\infty$, our reaction-diffusion 
equation for the concentration distribution $u_n(t)$ takes the form 
\begin{equation}
\frac{\partial u_n}{\partial t}  =
   F (u_n) 
   +D (u_{n+1}+u_{n-1}-2u_n) ,
  \label{bulkcell}  
\end{equation}
where $D$ is the diffusion constant, and $F(u)$ is a nonlinear
reaction kinetics function characterising the cells in the bulk. 
We assume that the
cell system is not coupled diffusively to its environment,
but by receptor molecules in the cell membrane, it is capable of
receiving an external inductive signal. Since real biological signal
transduction mechanisms are complicated cascades of different
enzime reactions, without worrying about the details here, we only assume
that due to the signal the reaction kinetics function in the cells
change. We consider the case when the penetration
depth of the signaling molecules is so short that the signal is
received almost exclusively by the very first cell along the line,
and, for simplicity, the signal is thought to effect the
morphogene production linearly.
(Note that we make a clear distinction between the signaling molecules
and the morphogene. The latter can freely diffuse in the system, while
the former cannot.)
With this proviso, we write the reaction-diffusion equation for
the {\em first} cell in the form
\begin{eqnarray}
\FL
\frac{\partial u_1}{\partial t}  = S +
  F(u_1)  + D(u_2-u_1).
  \label{firstcell}  
\end{eqnarray}

Out of the various theoretical possibilities, in the following we
analize the case when $F(u)$ is bistable and piece-wise linear
\begin{equation}
F(u)=\left\{ \begin{array}{ll}
              -\beta u      & \mbox{if $u< a$}\\
              -\beta (u-1)  & \mbox{if $u\ge a$}
              \end{array}
       \right.
\label{Flin}
\end{equation}
with $\beta>0$ and $0<a<1$. This is a charicature of the widely used Nagumo 
reaction kinetics function
\begin{equation}
F(u)=\beta u(1-u)(u-a).
\label{Fcont}
\end{equation}
In the sequal $\beta$ will be set $\beta=1$, which can always be achieved
by rescaling appropriately $t,D$ and $S$.

In both cases of $F(u)$ the reaction kinetics of the cell is 
{\em bistable} in the lack of the external
signal. When the signal is present, it acts as a chatalizer in cell 1 and
increases the production rate of the morphogene.
Even though the continuous form Eq.\ (\ref{Fcont}) is more realistic,
we study in detail
the piecewise linear charicature since it is analiticaly more tracktable.
Numerical simulations carried out using the Nagumo form Eq.\
(\ref{Fcont}) show that the qualitative behavior of the two models
are essentially the same. Some minor differences will be pointed out in 
the sequel.

In order to be able to assess the role of discreteness in the model we
will also consider its usual continuous space analog, i.e.,
when the discrete Laplacian is replaced by the second derivative
\begin{eqnarray}
&&n\to x \nonumber \\
&&\frac{\partial u}{\partial t}  =
D\frac{\partial^2 u}{\partial x^2} + F(u),\quad x\ge 0
\label{rd_cont}
\end{eqnarray}
Coupling to the environment via the $S$ term in Eqs.\ (\ref{firstcell})
translates into a Neumann boundary condition at the surface of the system
$\partial u(x,t)/\partial x|_{x=0}\sim S$.

The discrete and continuum models only become equivalent in the
large $D$ limit. This can
be easily shown by dimensional analysis. The only parameter whose
dimension contains the spatial length is D, $[D]=[m^2/s]$. As
$[\beta]=[1/s]$ any solution $u(x,t)$ of the 
continuum model must contain $x$ and $D$ in the
combination $x/\sqrt{D/\beta}$. When $D$ is large $u(x,t)$ varies slowly
in space so the second derivative can be discretized on the lattice
without committing much error. Note that the discrete version contains
an additional length scale: it is the lattice constant which was chosen
to be unity. The solution of the discrete model is expected to deviate
considerably from that of the continuous model
when the diffusion length
becomes comparable to the lattice constant, i.e., $\sqrt{D/\beta}\sim 1$.

The set of equations defined above contains the basic elements to
model cellular differentiation in response to an external signal:
Before switching on the inductive signal, our system is uniform 
(undifferentiated). The morphogene concentration in every cell is $u_n=0$, which
is clearly a stable steady state. We can say that all the cells have
type 0. When the external signal begins to increase (that we suppose
to be adiabatically slow), the first cell at the end of the chain goes 
through a bifurcation, and switches from the branch $u<a$
(type-0) to the branch $u>a$ (type-1). It becomes
differentiated. As the signal strength increases further, more and
more cells flip into type 1. This avalanche of
bifurcations may
become self-sustaining, and the differentiation may sweep through the
system in the form of a travelling wave. Under different conditions,
the position of the domain wall separating type-0 and type-1 cells
stays a well-defined function of the signal strength $S$. Then a
natural question is what happens when $S$ (adiabatically) returns to its
original zero value. (According to biological observations, inductive
signals are only present in a certain time interval of the
process of development.) As we will see soon, eventually the
already differentiated cells do not de-differentiate into type 0,
but maintain their type-1 state even in the lack of external signal.
The system shows hysteresis.

\section{Propagation failure}

We begin our analysis with the classification of the possible {\em
bulk} (i.e., far from the surface) behaviors. We analyze under what
conditions can a two-domain steady state solution exist,
when $u_n$ is a monotonic decaying function of the cell position
$n$ and $u_{-\infty}=1$, $u_\infty=0$. We suppose that the
domain wall (kink) is located between sites $M-1$ and $M$,
so that
\begin{equation}
\left. \begin{array}{ll}
              u^M_n \ge a  & \mbox{if $n\le M-1$}\\
              u^M_n <   a  & \mbox{if $n\ge M$,}
              \end{array}
       \right. 
\label{kink-M}
\end{equation}
where we introduced the superscript $M$ to explicitely denote the
position of the kink. A concentration distribution $u_n(t)$ satisfying
Eq.\ (\ref{kink-M}) will be called a kink-$M$.

It is well-known\cite{Aron} that in the continuum 
version of the model in Eq.\ (\ref{rd_cont}) for an infinite system ($S=0$),
a {\em steady state} kink can only exist in the
special case when $a=1/2$. If $a>1/2$, a kink-type initial
profile develops instead into a travelling wave in which
the domain wall travels with a constant speed $c$ leftward. On the
other hand, if $a<1/2$, the doman wall travels rightward.
%
%

In the lattice version of Eq.\ (\ref{bulkcell}), however, steady 
state domain wall solutions persist in a wide range of $a$ values. When
$a\in [u_-,u_+]$, with $u_\pm=u_\pm(D)$, the domain wall is 
{\em pinned\/} and its propagation is impeeded. This, so 
called, {\em propagation failure}\cite{pinning,Fath,Ern}
is due to spatial discreteness. Travelling wave behavior
only exits when $a <u_-$ or $a >u_+$. 

In the case of the piece-wise linear function $F(u)$ in 
Eq.\ (\ref{Flin}) the calculation of $u_-$ and $u_+$ is 
straightforward.\cite{Fath} A candidate kink-$M$ steady state 
solution of Eq.\ (\ref{bulkcell}) with $\partial u_n/\partial t=0$
can be looked for in the form
\begin{equation}
u^M_n =  \left\{ \begin{array}{ll}
              1+A e^{n\kappa} & \mbox{if $n\le M$}\\
              B e^{-n\kappa}  & \mbox{if $n> M$.}
              \end{array}
       \right. 
\label{bulkkink}
\end{equation}
Substituting this ansatz into Eq.\ (\ref{bulkcell}), the inverse
diffusion length $\kappa$ and the two constants $A$ and $B$ turn
out to be
\begin{eqnarray}
   \kappa &=& 2 \sinh^{-1}\sqrt{1/4D} \label{kappa}
\end{eqnarray}
and
\begin{eqnarray}
   A    &=&  -\frac{1}{e^\kappa+1} e^{-M\kappa}, \qquad
   B    =  \frac{e^\kappa}{e^\kappa+1} e^{M\kappa}
     \label{kappaAB}
\end{eqnarray}
Note however that when the ansatz in Eq.\ (\ref{bulkkink}) is used, one 
tacitly assumes that all cells on the left (right) of the kink are on 
the high (low) concentration branch of the piece-wise linear function
$F(u)$. Having found the solution in Eqs.\ (\ref{kappa},\ref{kappaAB}),
this assumption must be checked for consistency: The
concentration values obtained for the left ($L=M-1$) and right ($R=M$)
neighboring cells of the kink are
\begin{eqnarray}
   u^M_L     &=& \frac{1}{2}+\frac{1}{2}\tanh\frac{\kappa}{2} =
              \frac{1}{2}+\frac{1}{2}(1+4D)^{-1/2}
              \nonumber \\
   u^M_R &=& \frac{1}{2}-\frac{1}{2}\tanh\frac{\kappa}{2} =
              \frac{1}{2}-\frac{1}{2}(1+4D)^{-1/2},
\label{cM1}
\end{eqnarray}
thus the above calculation is only consistent if we find that
\begin{equation}
   u^M_L>a\quad {\rm and}\quad  u^M_R\le a.
   \label{consist}
\end{equation} 
This allows us to identify the pinning region boundaries for a given $D$ as 
\begin{equation}
   u_\pm = \frac{1}{2}\pm \frac{1}{2}(1+4D)^{-1/2}.
\end{equation}
The values of $u_+$ and $u_-$ are plotted in 
Fig.\ \ref{fig:bulk} as a function of the diffusion constant $D$.
Clearly, the obtained kink-$M$ steady state solution is only valid
in the shaded region of the diagram. On the other hand, when $(D,a)$ is
outside the shaded domain, the solution in
Eqs.\ (\ref{bulkkink})--(\ref{kappaAB}) is only a spurious solution.
In this region of $(D,a)$ there are no steady state
solutions; all initial conditions develop into travelling waves.


Even though the effect of lattice pinning is alike for the case of
the Nagumo-type (continuous) reaction function, exact calculation
of the pinning boundaries is not feasible. A perturbative approach in
the small $a$ limit was carried out in Ref.\ \cite{Ern}.
There is also an important difference how the wavefront speed $c$ scales
as, for a given $a$, the diffusion canstant approaches its
critical value $D_c=D_c(a)$. Simple bifurcation theory
analysis\cite{Ern} shows that in the continuous $F(u)$ case
$c$ scales following a power law with an exponent $1/2$, i.e.,
\begin{equation}
   c\sim (D-D_c)^{1/2}
\end{equation}
while for the piece-wise linear kinetics the singularity is
logarithmic
\begin{equation}
   c\sim -1/\ln(D-D_c).
\end{equation}
The latter form arises essentially from the nonanalicity (jump
discontinuity) of the $F(u)$ function in Eq.\ (\ref{Flin}), and
has been analysed in detail in Ref.\ \cite{Fath}.

\section{Induction with hysteresis}

Let us know investigate the inductive situation. As the signal $S$
increases the distribution of concentration values $u_n(t)$ in the
system becomes monotonically decreasing as a function of $n$. 
Even when the system is not in an equilibrium state we can define
an $M$ value characterizing the actual position of the domain wall 
separating type-1 and type-0 cells using
Eq.\ (\ref{kink-M}).
When seeking a {\em steady state} kink at site $M$, we solve 
the semi-infinite set of equations defined by Eqs.\ (\ref{bulkcell})
and (\ref{firstcell}) with $\partial u_n(t)/\partial t=0$.
Again, the analytic solution is attainable for the piece-wise
linearized kinetics. Working with the Ansatz
\begin{equation}
   u_n =  \left\{ \begin{array}{ll}
              1+ A e^{n\kappa}+B e^{-n\kappa} & \mbox{if $n\le M$}\\
              C e^{-n\kappa}  & \mbox{if $n\ge M+1$}
              \end{array}
       \right. ,
\label{surfkink}
\end{equation}
the unknown coefficients turn out to be 
\begin{eqnarray}
   A &=& -\frac{1}{e^\kappa+1}e^{-M\kappa} \\
   B &=& (e^\kappa-1)S -\frac{e^\kappa}{e^\kappa+1}e^{-M\kappa}  \\
   C &=& (e^\kappa-1)S +
     \frac{e^\kappa}{e^\kappa+1}(e^{M\kappa}-e^{-M\kappa})
\end{eqnarray}
with $\kappa$ again given by Eq.\ (\ref{kappa}). Unlike in the
translation invariant (infinite chain) case in Eq.\ (\ref{cM1}), 
the concentration values at the kink,
$u^M_L$ and $u^M_R$, are now explicite functions
of the kink position $M$ and the signal strength $S$
\begin{eqnarray}
   && u^M_L(S) = (e^\kappa-1)e^{-M\kappa} S
              +\frac{e^\kappa}{e^\kappa+1}(1-e^{-2M\kappa}) 
               \nonumber \\
   && u^M_R(S) = e^{-\kappa} u^M_L.
   \label{cM1S}
\end{eqnarray}
In the limit $M\to\infty$ we get back the bulk results in Eq.\
(\ref{cM1}).

As it was done for the infinite system, consistency of the solution must be
checked at this point. When the consistency condition of Eq.\ (\ref{consist})
fails no steady state solution with the kink at site $M$
exists. As a consequence the process of differentiation cannot stop at
site $M$, and the domain wall moves on.

Since the explicit expression
for $u^M_L(S)$ and $u^M_R(S)$ is available in Eq.\ (\ref{cM1S}), for any
values of $\beta,D,a$ and $S$ we can readily construct the set of 
possible $M$ values $\{M\}$ for which the consistency condition in 
Eq.\ (\ref{consist}) holds, and thus the kink-$M$ steady state exists. 
Although in theory every
element of this set $\{M\}$ could be realized as a steady state,
it is the previous history of the system (the initial conditions) and
the dynamics of the reaction-diffusion process which determines
which steady state (if any) gets finally
realized. This is in contrast with the continuum space model description
where the
position of a steady-state kink is always uniquely determined by the
actual model parameters.

The set of possible steady states on the $a$ vs $S$ plane 
for the case $\beta=1,D=2$ fixed is depicted in Fig.\ \ref{fig:steady}. The different
domains are separated by straight lines which is an artifact stemming from the
simple form of $F(u)$ in Eq.\ (\ref{Flin}). Nevertheless, a qualitatively
similar diagram can be obtained using the continuous Nagumo form. There
are three main possibilities for a given $a$ and $S$: (i) the number of
steady states is finite, (ii) any $M$ yields a valid steady state above
a certain value (this is indicated by a "+"), (iii) there are no steady
state kinks at all.

In the piece-wise linear model under investigation the kink steady
states, if they exist, are always stable against perturbations.\cite{Fath}
Thus if
at a given time $t$ the actual kink position is not an element of the
set of steady states $\{M\}$, differentiation or de-differentiation 
continues untill the
domain wall reaches the first $M$ value which is already in $\{M\}$. 
Having reached the domain of attraction of a stable steady state the kink
stabilizes at that point and the process halts untill, eventually, a further
change in $S$ destabilizes the system again.

Let us consider now an adiabatically slow process in which the inductive signal
increases from zero to $S_{\rm max}$ and then decreases back to zero again.
Using the above rule we can easily construct the phase diagram shown in
Fig.\ \ref{fig:phdiag} for such a process. Domains are labelled by the $M$
values of the kinks as they get realized in order. For example, the
small domain $[01230]$ has a history in which $M$ increases continuously
from 0 to 3, then as the signal diminishes it jumps ubruptly back to 0.
Once again there are three main regions, separated by thick lines, in this
phase diagram: \newline
(i) In the upper part of the phase diagram the system gradually
differentiates and then completely de-differentiates as
the signal varies. The de-differentiation process can be continuous or may
contain sudden jumps when the value of $a$ is closer to $u_+$. 
Note that this region corresponds more or less to the values of $a$
where in the infinite model kinks develop into travelling waves moving
leftwards, i.e. towards the surface of our semi-infinite system. 
Due to this bias a continuous presence of the signal is needed to
maintain differentiated type-1 cells in the system.\newline
(ii) In the middle part of the phase diagram, corresponding
approximatively to the
pinning region of Fig.\ \ref{fig:bulk}, the cells remain differentiated
even when $S$ falls back to zero. Note that when $a$ is close to $u_-$
the domain wall can take a huge jump at the beginning as $S$ reaches a
certain value. In this situation the maximum value of the signal $S_{\rm
max}$ is important, since this is the factor which determines the range 
of the irreversibly differentiated domain. \newline
(iii) Finally for small values of $a$ the differentiation process 
becomes self-sustaining when a critical value of the signal is exceeded.
This mimics the infinite chain behaviour with a travelling wave moving 
rightwards. The signal only triggers the differentiation process but
after that it plays no further role.\newline
Typical examples of the three kinds of behavior are depicted
in Fig.\ \ref{fig:hister}(a-c). 

The structure of the phase diagram in Fig.\ \ref{fig:phdiag} is
rather involved, demonstrating that even a simple model like this can show
an amazing complexity. When the more realistic Nagumo-type reaction
function is considered the exact solvability of the problem is lost.
Nevertheless, numerical simulations we carried out demonstrated that the
main conclusions about the qualitative behavior of the three phases 
remain unchanged. 

\section{Summary and discussion}

In this paper we analysed a semi-infinite one-dimensional
one-chemical reaction-diffusion
system. The reaction kinetics was assumed to be bistable, giving
rise to two different type of cells: type-0 (low concentration type)
and type-1 (high concentration type). Starting from a homogeneous
situation (all cell are type-0) the system underwent a 
differentiation process in response to an external inductive signal.
The signal was modelled as a boundary condition in the continuum version
and as an extra term in the internal cell kinetics of the first cell
in the discrete space version.

Depending on the model parameters the
differentiation process sweeps through the whole system, or
flips a limited number of cells to type-1 up to a given position. We
found that the behavior of the system differs considerably in
the continuum and in the discrete space versions. In the former the
position of the domain wall between the two cell types is either a
well-defined function of the external signal strength, or the front of
differentiation inevitably develops into a travelling wave. In the
discrete case the fate of the system depends on its previous history,
giving rise to hysteresis.   

We analysed in detail the situation when the
bistable reaction function is piece-wise linear.  
The model was solved analytically, and we constructed a
detailed phase diagram based on the different types of behavior
as a function of the model parameters.
We found three mayor scenarios for the system: (i) In response to the
inductive signal the solution develops into a travelling wave which
differentiates the whole (semi-infinite) domain, (ii) the signal causes
some spatially limited differentiation but when it diminishes all
cells de-differentiate, (iii) differentiated cells get stabilized
and the inhomogenious solution persists even when the signal disappears.

Although the analysis was carried out with a somewhat special reaction
function, numerical simulations we have done support our expectation that
the observed behavior is widely universal in discrete space models, and 
the qualitative chategories found remain valid in similar models with more 
realistic reaction functions. 
There are, of course, minor quantitative differences such as the type of
scaling near the bifurcation points, or the actual domain wall location for
a given signaling scheme.

In general we have found that adding spatial discreteness to reaction
diffusion models has a tendency to improve domain wall stability between
different tissues, and to make the emerging pattern less
susceptible to fluctuations of the signaling mechanisms.
Since robostness and stability of 
developmental processes and that of the adult organism is an inevitable 
necessity for the survival of biological species, we may wonder that 
the invention of cellular membranes by evolution, which made the
fundamental building blocks discrete, was at least in parts 
motivated by such a developmental benefit, among others.

Finally we would like to mention an actual biological observation
which seems to be explainable on the basis of the above model.
During the retina differentiation of Drosophila it has been
observed that ommatidia (the basic functional units of the retina
consisted of photoreceptor and other types of cells) develop behind a 
slowly moving wave front, the "morphogenetic furrow".\cite{retina} There
are many genes which are only expressed behind the furrow, and thus one 
(or more) of them is believed to play the role of
a morphogene. It was also noticed\cite{Koc-Mei} that a slight shift in the
environmental temperature is enough to make the wave front stop. Untill
the temperature is raised back to normal again the process of 
differentiation does not continue. This slight artificial manipulation,
although capable of empeeding the propagation of the front for hours or
days, is believed to have no residual effects on further retina development.

Knowing that the propagating wavefront is extremely slow, $c\approx ?
m/s$, we can speculate that the developing retina is in fact tuned very
close to a pinning region boundary. A change in the tissue temperature
necessarily alters the actual model parameters. Although it would be
very difficult to estimate on a phenomenological basis how the complex, 
non-linear set of biochemical reactions get modified by a temperature 
decrese, we can at least assume that the overall diffusivity of the
chemicals get reduced. This can drive the system into
the pinning region as is illustrated in Fig.\ {fig:bulk}, causing eventually
a propagation failure. In this unfavorable temperature regime the domain
wall, represented by the morphogenetic furrow, between undifferentiated
and already differentiated cells becomes a stable steady state, and the
differentiation process temporarily halts.

Valuable discussions with Paul Erd\H os are acknowledged.
This research was supported by the Swiss National Science Foundation 
Grant No.\ 20-37642.93.

\begin{widetext}

\begin{figure}[tbp]
\includegraphics[scale=.7]{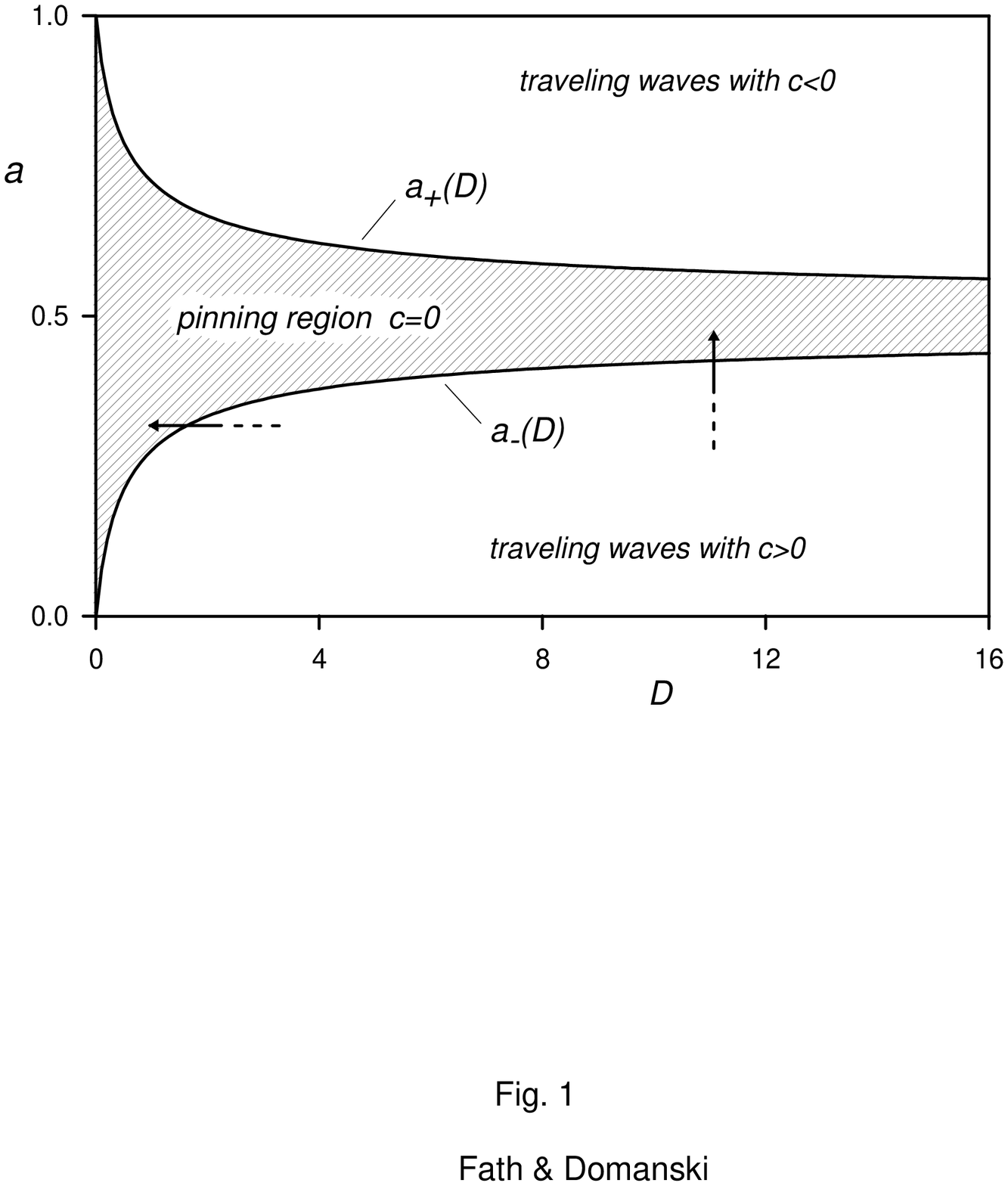} 
\caption{Phase diagram on the $a$ vs $D$ plane for an infinite chain. 
Pinned steady state
solutions exist in the shaded region, while traveling waves exist
in the unshaded ones. The pinning transition takes place along
the $a=a_+(D)$ and the $a=a_-(D)$ curves, and can be initiated either by
changing $D$ for $a$ fixed, or by varying $a$ with $D$ fixed (see
arrows).} 
\label{fig:bulk} 
\end{figure}

\begin{figure}[tbp]
\includegraphics[scale=.7]{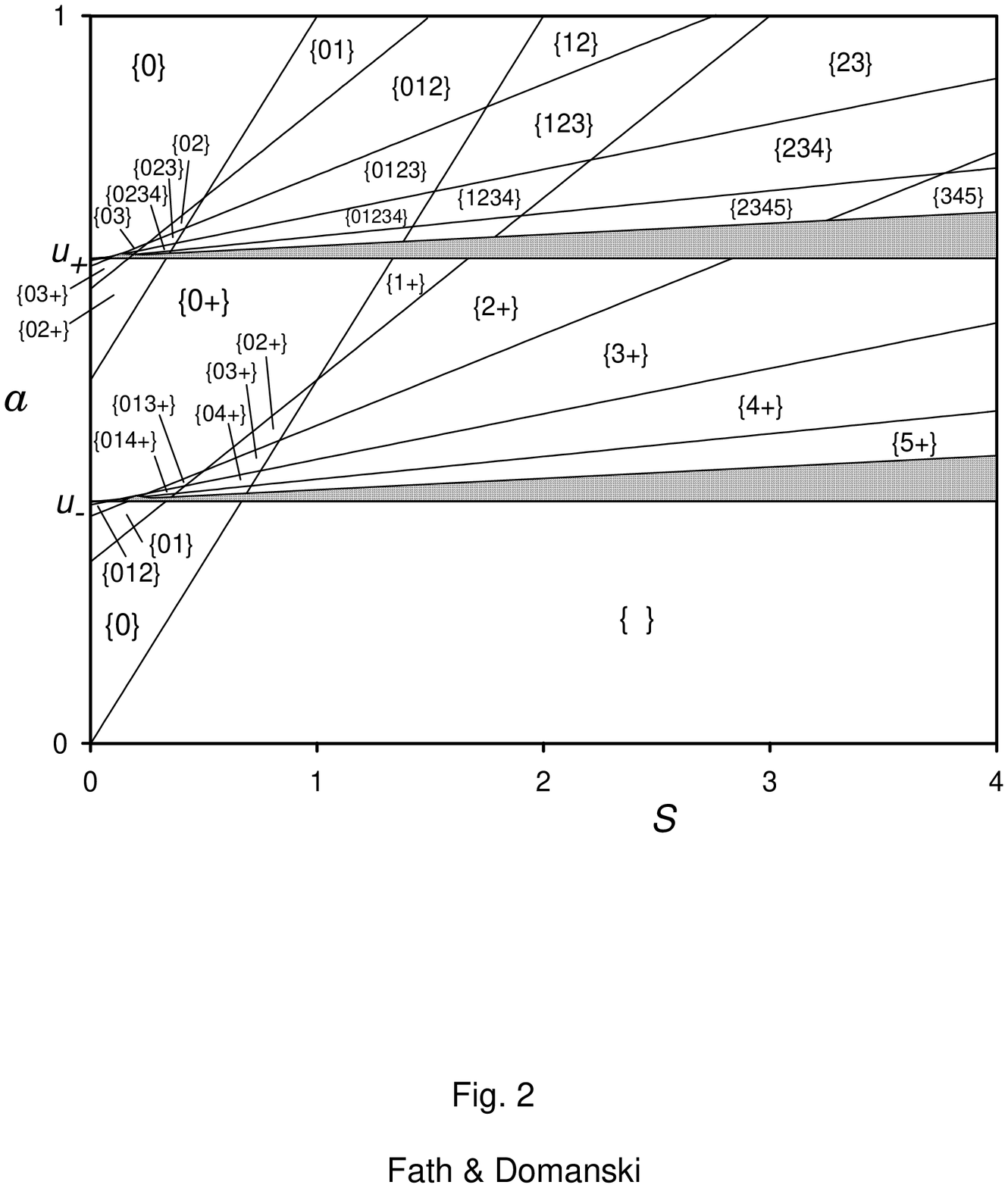} 
\caption{Steady state kink solutions on the $a$ vs $S$ plane with $D=2$. The set
$\{M_1M_2\cdots\}$ denotes the possible positions of the kinks in the given region.
A "+" represents that all $M$ values are possible above the preceding
value. In the shaded regimes many small phases appear.} 
\label{fig:steady} 
\end{figure}

\begin{figure}[tbp]
\includegraphics[scale=.7]{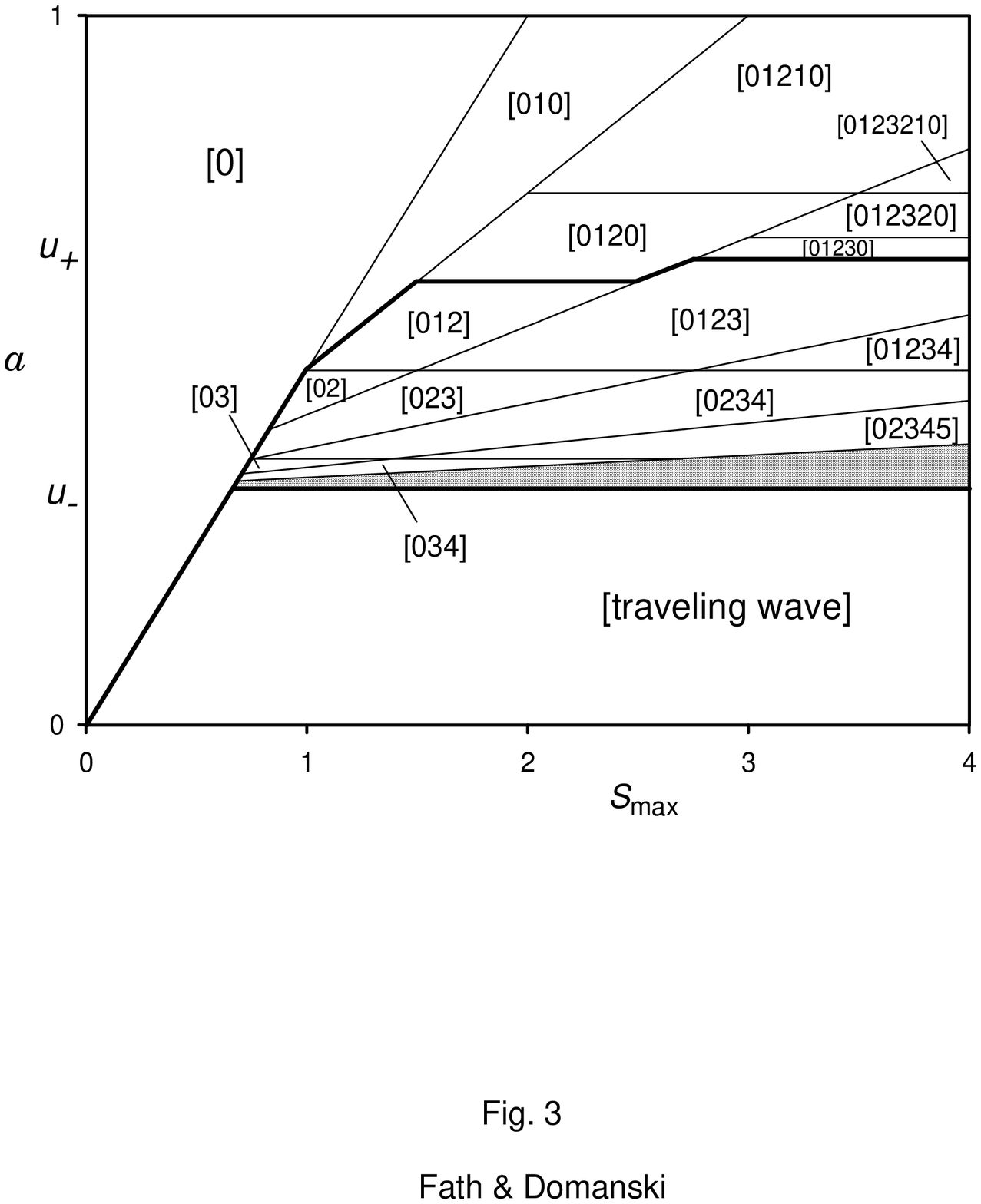} 
\caption{Phase diagram for the signaling scheme $S:\;0\to S_{\rm max}\to
0$ on the $a$ vs $S_{\rm max}$ plane with $D=2$. The symbol
$[M_1M_2\cdots]$ denotes the order of kink positions as they get
realized. In the shaded regime many small phases appear.} 
\label{fig:phdiag} 
\end{figure}

\begin{figure}[tbp]
\includegraphics[scale=.7]{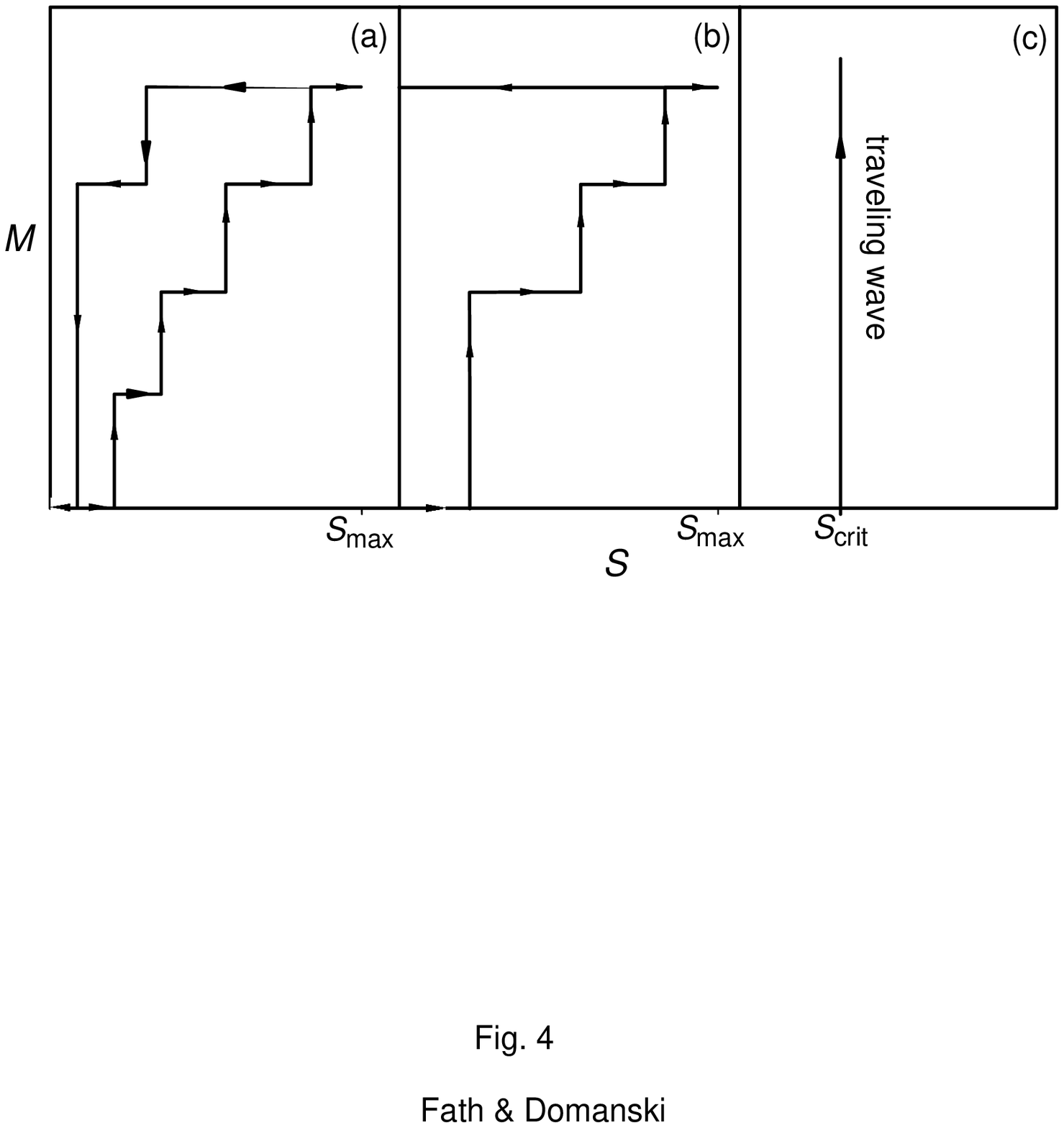} 
\caption{Schematic hysteresis diagrams for the signaling scheme 
$S:\;0\to S_{\rm max}\to 0$ showing the actual domain wall position $M$
as a function of the signal strength $S$. (a) all cells de-differentiate,
(b) some cells remain differentiated, (c) all cells become differentiated as
a travelling wave emerges.} 
\label{fig:hister} 
\end{figure}

\end{widetext}

\end{document}